\documentclass[11pt,a4paper]{article}
\usepackage{jheppub}
\usepackage{epsf}
\usepackage{amsmath}
\usepackage{amssymb}
\usepackage{amsfonts}
\usepackage{graphicx}
\usepackage{graphics}
\usepackage{epstopdf}
\usepackage{url}

\newcommand{\beq}{\begin{equation}}
\newcommand{\eeq}{\end{equation}} 
\newcommand{\beqa}{\begin{eqnarray}} 
\newcommand{\eeqa}{\end{eqnarray}}

\def\bi{\begin{itemize}}
\def\ei{\end{itemize}}
\def\be{\begin{equation}}
 \def\ee{\end{equation}}
\def\ben{\begin{equation*}}
 \def\een{\end{equation*}}
 \def\bea{\begin{eqnarray}}
 \def\eea{\end{eqnarray}}
 \def\bean{\begin{eqnarray*}}
 \def\eean{\end{eqnarray*}}

\newcommand{\ed}{\end{document}}

\newcommand{\jetphox}{\texttt{JETPHOX}}

\newcommand{\zh}{z_{_{\rm h}}}
\newcommand{\pt}{p_{_\perp}}
\newcommand{\ptjet}{p_{_\perp}^{\rm jet}}
\newcommand{\pth}{p_{_\perp}^{\rm h}}

\newcommand{\dd}{{\rm d}}
  
 \newcommand{\gsim}{\gtrsim}

  \def\pt{p_{\perp}}

  \def\kt{{k_{\perp}}}

 \def\esim{\,\mathrel{\rlap{\lower0.2em\hbox{$-$}}\raise0.15em\hbox{\footnotesize $\hskip0.04em\sim$}}\,}
 \def\gsim{\mathrel{\rlap{\lower0.2em\hbox{$\sim$}}\raise0.2em\hbox{$>$}}}
 \def\ksim{\mathrel{\rlap{\lower0.2em\hbox{$\sim$}}\raise0.2em\hbox{$<$}}}

\def\epem{$e^+e^-$}

\title{Probing fragmentation functions from same-side hadron--jet momentum correlations in p--p collisions}

\author[a,b]{Fran\c{c}ois Arleo,}
\author[c]{Michel Fontannaz,}
\author[a]{Jean-Philippe Guillet,}
\author[a]{Chi Linh Nguyen}

\affiliation[a]{Laboratoire d'Annecy-le-Vieux de Physique Th\'eorique (LAPTh)\\ UMR5108, Universit\'e de Savoie, CNRS, BP 110, 74941 Annecy-le-Vieux cedex, France}
\affiliation[b]{Laboratoire Leprince-Ringuet (LLR)\\ UMR 7638,  \'Ecole polytechnique, CNRS, 91128 Palaiseau, France}
\affiliation[c]{Laboratoire de Physique Th\'eorique (LPT)\\
UMR 8627, Universit\'e de Paris XI, CNRS, 91405 Orsay cedex, France}

\emailAdd{arleo@lapth.cnrs.fr}
\emailAdd{michel.fontannaz@th.u-psud.fr}
\emailAdd{guillet@lapth.cnrs.fr}
\emailAdd{chi-linh.nguyen@lapth.cnrs.fr}

\abstract{A next-to-leading order (NLO) analysis of hadron--jet momentum correlations in p--p collisions at the LHC is carried out. We show that the inclusive charged hadron momentum distributions inside jets is a very sensitive observable which allows one to disentangle among various fragmentation function sets presently available. Correlations with identified hadrons (kaons, protons) are investigated as well.}


\begin{document} 

\maketitle
\setcounter{footnote}{0}
\renewcommand{\thefootnote}{\arabic{footnote}} 	


\section{Introduction}

The perturbative QCD calculation of large transverse-momentum ($\pt$) hadron production at hadronic colliders requires the knowledge of the non-perturbative fragmentation functions (FF), $D_i^h$, which describe the transition from partons to hadrons. On top of constraining non-perturbative aspects of QCD, fragmentation functions are also often used and needed in the context of ``jet quenching'' studies in heavy-ion collisions, in order to describe parton energy loss processes in quark-gluon plasma (see e.g.~\cite{Arleo:2008dn}).

Fragmentation functions have first been determined from global fits of \epem\ data (e.g. BFGW~\cite{Bourhis:2000gs}, KKP~\cite{Kniehl:2000fe}, Kretzer~\cite{Kretzer:2000yf}), at LEP and other facilities at lower energies. However, measurements in $e^+e^-$ collisions essentially constrain the \emph{quark} FF\footnote{In \epem\ collisions, gluon FF, $D_g^h$, can only be probed via scaling violations of $D_q^h$, or through 3-jet events.} and at not too large momentum fraction $z$. In order to get additional constraints, various groups recently included data on single hadron production at hadronic colliders, e.g. RHIC (AKK08~\cite{Albino:2008fy}, DSS~\cite{deFlorian:2007hc,deFlorian:2007aj}), Tevatron (AKK08), as well as data in low-$Q^2$ semi-inclusive deep inelastic scattering (DSS). Also, attempts to estimate theoretical uncertainties have been performed by DSS and HKNS~\cite{Hirai:2007cx} which confirmed the lack of constraints on gluon FF at large $z$.

Unlike in \epem\ collisions, for which the (anti-)quark momentum is known at leading order, single hadron production in hadronic collisions does not allow the energy of the fragmenting parton to be estimated. As a consequence, the measurement of hadron $\pt$-spectra in p--p collisions is sensitive to some moments of FF only.\footnote{A recent analysis~\cite{dEnterria:2013vba} of collider data compared with theoretical predictions based on these FF
parametrizations lead to the conclusion that most of the theoretical predictions tend to
overpredict the measured LHC and Tevatron cross sections.} On the contrary, performing momentum correlations in double inclusive hadron--jet production would in principle be able to probe more precisely the $z$ dependence of fragmentation functions. 
Similarly, analyses of photon--jet~\cite{Belghobsi:2009hx} and photon--hadron~\cite{deFlorian:2010vy} momentum correlations aiming at setting additional constraints on FF into photons and into hadrons have also been carried out recently. In the latter study, the photon is produced in the away side of the measured hadron; its momentum can therefore serve as a proxy for that of the recoiling parton as long as only one jet is produced in the event, i.e. if real higher order corrections (with $2\to3$ kinematics for the parton scattering dynamics) remain small. In order to circumvent this issue and to increase counting rates, we investigate in this study  the energy distribution of energetic hadrons inside identified jets in p--p collisions at the LHC as a mean to further constrain FF. The analysis is carried out at NLO accuracy with \jetphox\ and using various FF sets available (AKK08, BFGW, DSS, HKNS, Kretzer). On the experimental side, such a measurement has been measured e.g. by CMS in Ref.~\cite{Chatrchyan:2012gw} although this study focused on the medium-modifications of hadron distributions at small energy fraction, $z\ll 1$.

The outline of the paper is as follows. In section~\ref{se:theory} we present and motivate the theoretical framework of this study. Results on charged and identified hadron--jet momentum correlations are shown respectively in section~\ref{se:chargedhadrons} and section~\ref{se:identifiedhadrons} and discussed. Finally we conclude in section~\ref{se:conclusion}.

\section{Framework}\label{se:theory}

\subsection{Perturbative calculation}\label{se:pqcd}

The double-inclusive hadron--jet production cross section
is computed in p--p collisions at $\sqrt{s}=8$~TeV at next-to-leading order (NLO) accuracy using the \jetphox\ Monte Carlo program~\cite{Catani:2002ny}, with CTEQ6.6~\cite{Nadolsky:2008zw} parton distribution functions. Jets are reconstructed using the $\kt$ algorithm\footnote{Note that at NLO accuracy, there are at most two partons in a given hemisphere, making in this context the widely used anti-$\kt$ algorithm~\cite{Cacciari:2008gp} exactly equivalent to the $\kt$ algorithm.}~\cite{Catani:1993hr,Ellis:1993tq} with a jet radius $R=0.4$.

The initial-state factorization scale, $M$, and the renormalization scale, $\mu$, are taken to be equal and proportional to the jet transverse momentum, $M=\mu=  \ptjet$. For the fragmentation scale we use $M_F = R \  \ptjet$ in order to resum in the fragmentation function $D_q^h (z, M_F)$ all the $\log\left(R\ \ptjet / M_F\right)$-terms present in the higher-order corrections.\footnote{A detailed discussion of the resummation of the $\log\left(R\ \ptjet / M_F\right)$-terms is given in Ref.~\cite{Catani:2013oma}.} 

In order to estimate the uncertainty of the NLO predictions, all scales are varied simultaneously by a factor of two, up and down, in the calculations. The scale dependence of our results will be discussed in more detail in section~\ref{se:chargedhadrons} and section~\ref{se:identifiedhadrons}.

\subsection{Fragmentation function sets}\label{se:ffsets}

\begin{figure}[htb]
\begin{center}
    \includegraphics[width=7.45cm]{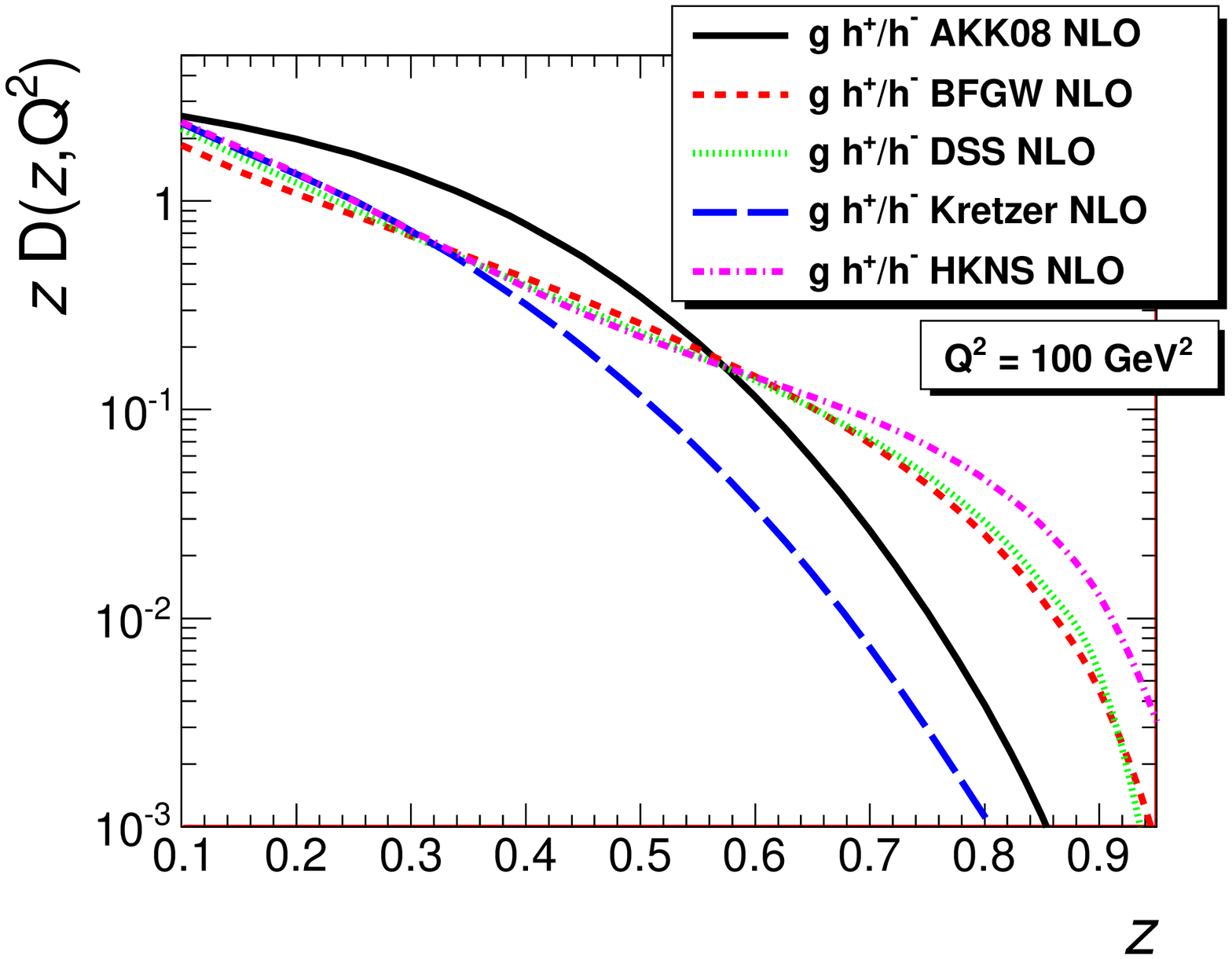}
    \includegraphics[width=7.45cm]{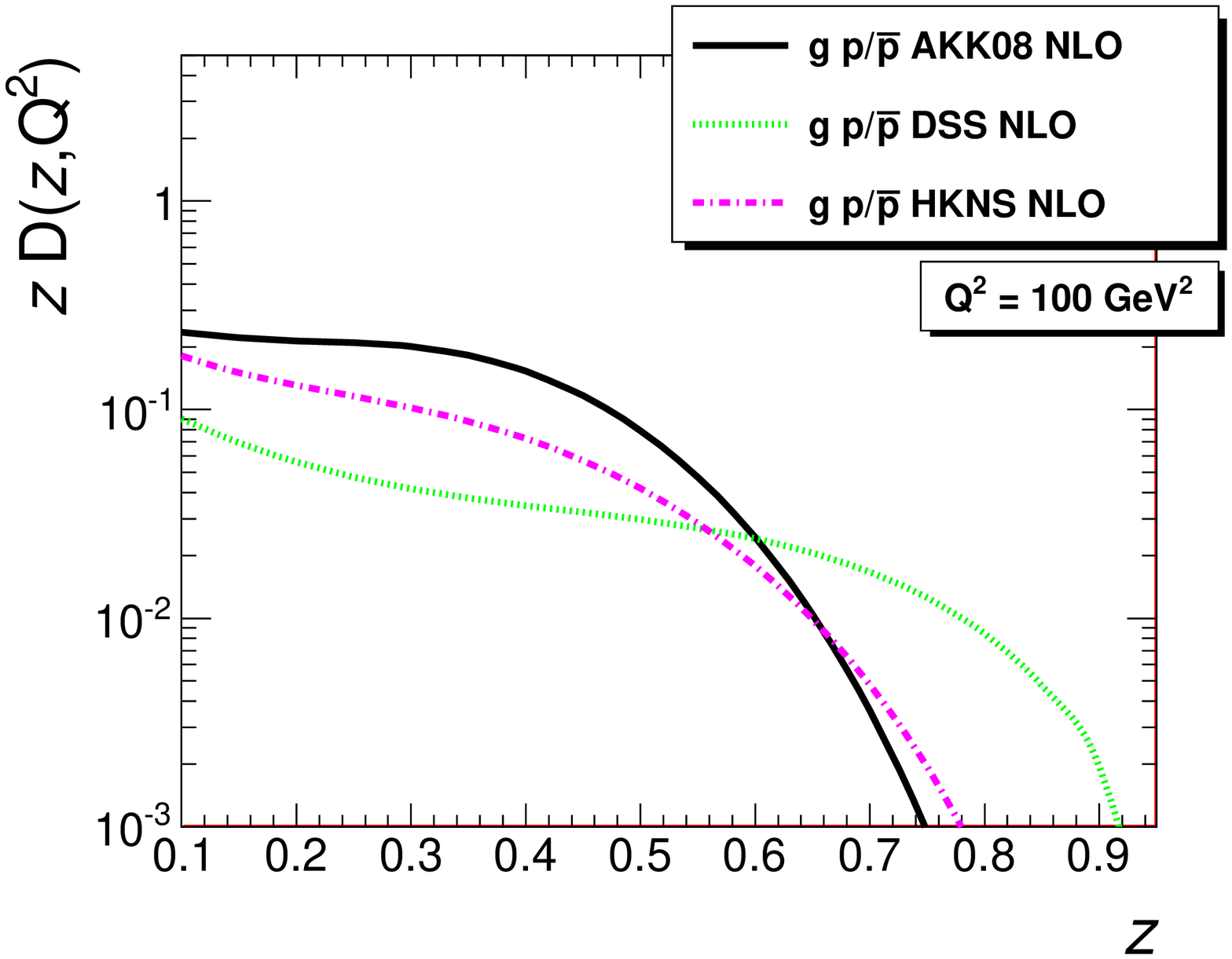}
  \end{center}
\vspace{-1cm}
\caption{Comparison of  gluon fragmentation functions into charged hadrons (left) and protons+anti-protons (right), as a function of $z$ and at $Q^2=100$~GeV$^2$.}
  \label{fig:ffgenerator}
\end{figure}

The goal of the paper is to explore the sensitivity of fragmentation functions on the jet--hadron momentum correlations. We shall therefore compare results using various FF sets presently available, namely AKK08~\cite{Albino:2008fy}, BFGW~\cite{Bourhis:2000gs}, DSS~\cite{deFlorian:2007hc}, HKNS~\cite{Hirai:2007cx}, Kretzer~\cite{Kretzer:2000yf}. 
(In addition it is also possible in principle, yet CPU-time expensive, to perform the NLO calculations using the theoretical uncertainty bands provided by the DSS and HKNS sets. This goes beyond the scope of this prospective study and is left for future work when precise data become available.)
This choice reflects well the variety of the FF sets and the spread in the different predictions, both in shape and in magnitude. In order to illustrate this, the gluon fragmentation into charged hadrons is plotted\footnote{We used the FFGenerator \url{http://lapth.cnrs.fr/ffgenerator}.} in Fig.~\ref{fig:ffgenerator} (left) as a function of $z$ ($Q^2=100$~GeV$^2$), showing significant differences between the different parametrizations, especially at large values of $z$. As shown later, the hadron--jet momentum correlations exhibit similar features thus allowing one to disentangle among the various sets available. The spread among the different FF become even larger when considering fragmentation into identified hadrons, as shown from the AKK08, DSS and HKNS parametrizations of FF into protons+anti-protons (Fig.~\ref{fig:ffgenerator}, right).

\subsection{Hadron momentum distribution inside jets}\label{se:variables}

We consider the distribution in the momentum fraction
\begin{equation}\label{eq:imbalance}
\zh =  {\vec{p}_{\perp}^{{\rm h}} \cdot
\vec{p}_{\perp}^{\,\rm{jet}} \over |\!|\vec{p}_{\perp}^{\,\rm{jet}}|\!|^2}
\end{equation}
carried by hadrons inside identified jets. At leading order, the fraction $\zh$ reduces to the fragmentation variable $z$. Also note that the typical angle between the hadron and the jet direction is very small, therefore almost identical results would be obtained using transverse momentum, ${{p}_{\perp}^{{\rm h}}} /
{p}_{\perp}^{\,\rm{jet}}$, or energy, ${{E}^{{\rm h}}} /
{E}^{\,\rm{jet}}$, fractions. For consistency with the photon--jet analysis~\cite{Belghobsi:2009hx}, we shall keep the usual momentum imbalance variable (\ref{eq:imbalance}).

The distribution $\dd\sigma/\dd \zh$ of hadrons inside jets of \emph{fixed} momentum $\ptjet$ should therefore directly reflect the $z$ dependence of the fragmentation functions at a hard scale $Q \sim R\ {\ptjet}$. 
In this paper, we rather 
propose to study the $\zh$ distribution of hadrons inside jets of \emph{all} transverse momenta above $\pth$, i.e. $\pth < \ptjet <\sqrt{s}/2$. With such a requirement, the distribution reflects, at leading order, the (un-normalized) conditional probability distribution that the hadron of momentum $\pth$ carries the momentum fraction $\zh$ of its parent parton. As a consequence, these distributions are naturally peaked at large values of $\zh \gtrsim 0.5$, i.e. the relevant range for hadron production in hadronic collisions, unlike the distribution of hadrons produced in \epem\ collisions.

\subsection{Cuts}

Because of the QCD evolution, differences between the various FF sets are expected to weaken at very large scales, $Q \gg \Lambda_{\rm QCD}$, since this evolution is only logarithmic.
Also note that at very large $\ptjet$,
 hadrons predominantly come from the fragmentation of quarks instead of gluons, for which FF are better constrained from \epem\ measurements (see below section~\ref{se:flavour}).
Therefore, in order to possibly disentangle (gluon) fragmentation functions from hadron--jet momentum correlations we require the jet momentum to be not too large, nevertheless keeping in mind that the experimental jet reconstruction cannot be achieved below a given transverse momentum. We apply in this analysis a minimal hadron/jet $\pt$-cut of $\left(\ptjet\right)^{\rm min} = \left(\pth\right)^{\rm min}=30$~GeV above which the experimental determination of the jet energy scale remains under control. We also restrict the hadron transverse momenta to remain below\footnote{This cut will have basically no effect on the computed distributions; however this avoids using FF at very large scales for which they are not always available.} $\pth< \left(\pth\right)^{\rm max}=200$~GeV and apply no restriction on the jet upper transverse momentum. Finally, a lower cut on the momentum imbalance is applied, $\zh > 0.1$, below which the fixed-order calculation may no longer be appropriate because of the appearance of large logarithms $\ln(1/\zh)$ which would need to be resummed to all orders.

\subsection{Flavour composition}\label{se:flavour}

As above mentioned, hadron production in \epem\ collisions naturally comes predominantly from the fragmentation of quarks and anti-quarks. In high-energy p--p collisions, however, gluon production dominate over quark production at small $\ptjet/\sqrt{s}$. As a consequence, hadron production essentially arises from gluon fragmentation, at least on a very large range of momentum fractions carried by the detected hadron. It is one of the reasons why the theoretical uncertainty associated to the calculation of large-$\pt$ hadron production at LHC is important because of the rather unknown gluon FF~\cite{dEnterria:2013vba,Arleo:2010kw}.

\begin{figure}[htb]
\begin{center}
    \includegraphics[width=14cm]{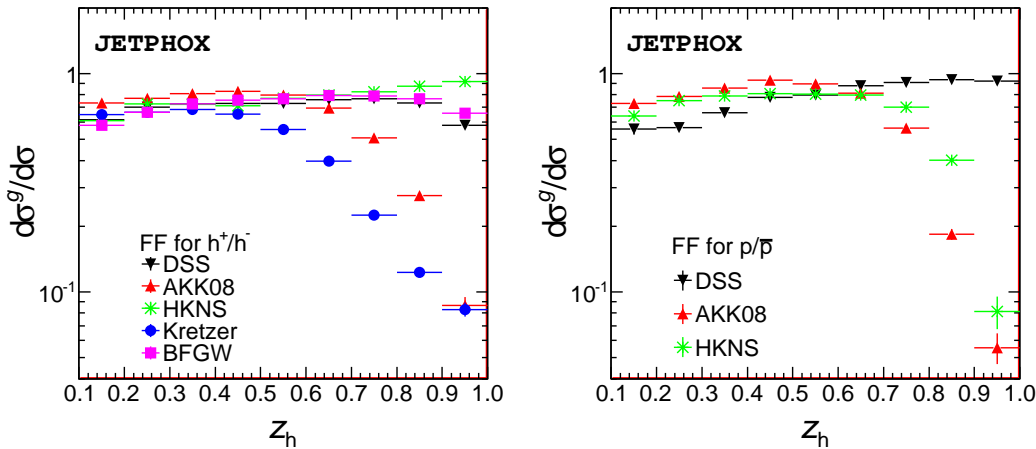}
  \end{center}
\vspace{-1cm}
\caption{Relative contribution of gluon fragmentation to the production of charged hadrons (left) and protons+antiprotons (right) inside jets, using the various FF sets. See text for details.}
  \label{fig:flavour}
\end{figure}

In order to illustrate this, $\zh$ distributions have been computed using the different sets assuming gluon fragmentation only (by setting artificially the quark fragmentation to zero), normalized to the ``full'' $\zh$ distribution, i.e. including both quark and gluon fragmentation. The result is shown in Fig.~\ref{fig:flavour} for the different FF sets and for both charged hadron (left) and protons+antiprotons (right) production. As can be seen, Fig.~\ref{fig:flavour} confirms the dominance of gluon fragmentation, which contributes to 60-80\% to charged hadron production at all $\zh$ for the BFGW, DSS and HKNS sets. Interestingly, charged hadron production proceeds essentially through quark fragmentation above $\zh \gtrsim 0.75$ (respectively $\zh \gtrsim 0.6$) when using the AKK08 (respectively, Kretzer) FF set; the reason comes from the very soft gluon fragmentation function of these two sets, see Fig.~\ref{fig:ffgenerator}. Regarding protons+antiprotons production, the situation is analogous for the DSS and AKK08 sets. The HKNS fragmentation function, however, leads to a strong depletion of gluon to (anti)protons at large $\zh$, and similar to AKK08, unlike what was observed for charged hadron production. This could also have been anticipated from a glance at Fig.~\ref{fig:ffgenerator}.

\section{Results}

\subsection{Correlations with inclusive charged hadrons}\label{se:chargedhadrons}

\begin{figure}[htb]
\begin{center}
    \includegraphics[width=14cm]{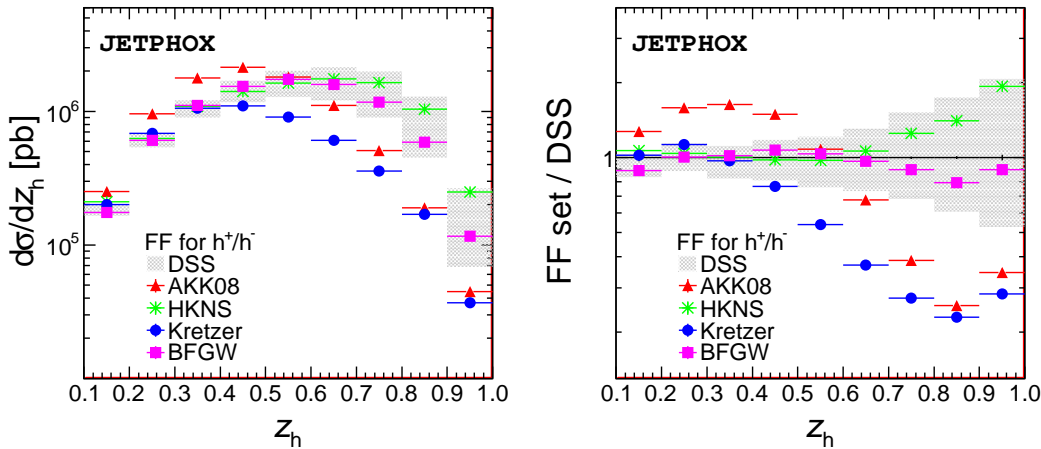}
  \end{center}
\caption{Left: $\zh$ distributions of charged hadrons inside jets, using the AKK08, BFGW, DSS, HKNS and Kretzer FF sets. The band indicates the scale dependence of the DSS calculation (see text). Right: Same distributions normalized to the DSS prediction.}
  \label{fig:zabs}
\end{figure}

The $\zh$ distribution of inclusive charged hadrons in p--p collisions at the LHC (we choose $\sqrt{s}=8$~TeV) is shown in Fig.~\ref{fig:zabs} (left) using the AKK08, BFGW, DSS, HKNS and Kretzer FF sets and the scales given by the central values discussed in section~\ref{se:pqcd}.

 As can be seen, differences between the various predictions can be rather large. The distributions using the BFGW and DSS (and to a lesser extent HKNS) fragmentation functions prove rather similar, both in shape and in magnitude. Distributions using AKK08 and Kretzer have a similar shape --~yet a different magnitude~-- and somehow steeper than the results obtained using BFGW, DSS and HKNS. Such features are reminiscent to those of the gluon fragmentation functions (Fig.~\ref{fig:ffgenerator}) which appeared significantly softer for AKK08 and Kretzer; it is a hint that this observable should provide tight constraints on the various FF sets.
  
Another interesting observation is the scale dependence of the predictions, which quantifies the strength of higher-order corrections, shown as a band in the DSS prediction in Fig.~\ref{fig:zabs} (left). Although the scale dependence is not negligible at large $\zh$, it is remarkable that the spread of the predictions using the various FF sets exceed somehow the scale dependence of the calculations. In other words, the discrepancy between the different sets proves beyond the intrinsic uncertainty of the NLO predictions.

In order to be more quantitative, the various NLO calculations are normalized to those using a set of reference (here taken to be DSS), $\dd\sigma^{{\rm FF\ set}}\big/ \dd\sigma^{\rm DSS}$, see Fig.~\ref{fig:zabs} (right). This figure illustrates  further the different shapes expected when using BFGW and DSS on the one hand and AKK08 and Kretzer on the other hand. The shape of HKNS is rather similar to that of BFGW/DSS except at large $\zh \gtrsim 0.7$. The ratio $r$ between the FF sets can be significant at large $\zh$, from $r=0.3$ (Kretzer/DSS) to $r=1.4$ (HKNS/DSS) at $\zh=0.8$.

 As already noted, the scale dependence becomes increasingly large as $\zh$ gets closer to 1, from $10\%$ at $\zh\simeq 0.1$ up to $50$--$80\%$ at $\zh\gtrsim 0.8$. The origin is twofold. At large $\zh$, the extra radiated parton in $2\to3$ processes is forced to be soft, leading to large logarithms $\ln^2(1-\zh)$  which would need to be resummed to all orders (a work which is beyond the scope of the present paper). As a consequence, the scale variation at NLO becomes of the same order as the one at leading order when $\zh$ is close to 1. The other reason comes from the behavior of the anomalous dimension of the fragmentation functions, $\mu^2 \partial \ln D(z,\mu^2)/ \partial \mu^2$,  which becomes increasingly large as $z\to1$. Note however that the scale dependence of the present NLO calculation remains below the spread of the various calculations, at least when comparing AKK08 \& Kretzer to BFGW \& DSS.

The differences between the FF sets prove largest at very large $\zh$ (say, $\zh\gtrsim0.8$), where the differential cross section becomes very much suppressed. However, note that with the cuts used in this analysis, the counting rates remain significant even in the highest $\zh$ bin, thanks to the huge integrated luminosity delivered at the LHC. Taking ${\cal L}=20$~fb$^{-1}$ at $\sqrt{s}=8$~TeV~\cite{luminosityatlas,luminositycms}, the expected rates in the bin $\zh=[0.9,1.0]$ are ${\cal N}=7\times10^4$ using the lowest prediction $\dd\sigma/\dd\zh=3.5$~pb given by the Kretzer set.

\begin{figure}[htb]
\begin{center}
    \includegraphics[width=14cm]{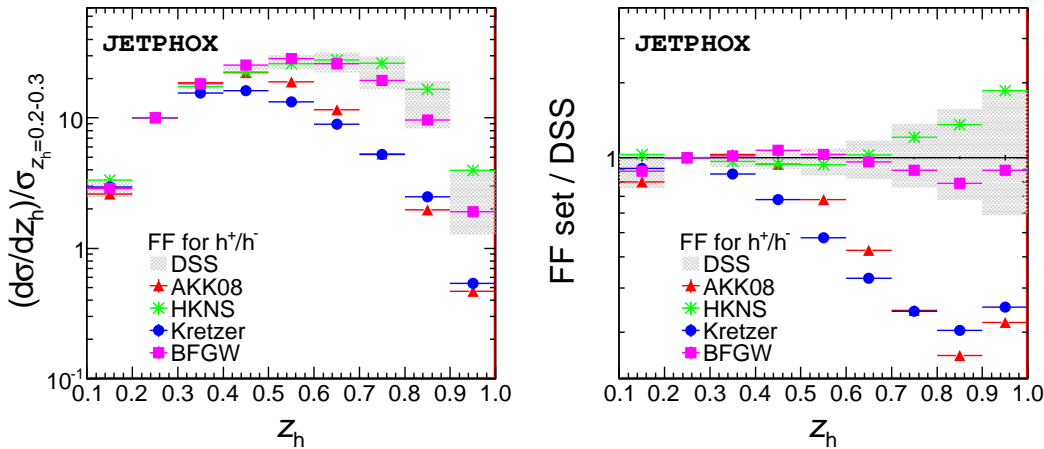}
  \end{center}
\caption{Left: Normalized $\zh$ distributions of charged hadrons inside jets, using the AKK08, BFGW, DSS, HKNS and Kretzer FF sets. The band indicates the scale dependence of the DSS calculation (see text). Right: Same distributions normalized to the DSS prediction.}
  \label{fig:znorm}
\end{figure}

Despite very different shapes, it might be difficult to disentangle, say, AKK08 from BFGW/DSS predictions, from the \emph{absolute} distribution $\dd\sigma/\dd\zh$. In order to truly probe the \emph{shape} of the $\zh$ distribution, we determine the distribution normalized to its value in the bin $\zh=[0.2,0.3]$, $\left(\dd\sigma/\dd\zh\right) \big/ {\left(\dd\sigma/\dd\zh\right)}_{\zh=[0.2,0.3]}$ in Fig.~\ref{fig:znorm} (left). Results using AKK08 and Kretzer FF prove rather similar since both predictions on  the absolute $\dd\sigma/\dd\zh$ essentially differ in the overall magnitude (see Fig.~\ref{fig:zabs}), which cancels in the normalized distributions. Apart from emphasizing the shape of the FF, a clear advantage of the normalized distribution is to reduce the scale dependence of the NLO calculations, which mostly affects the magnitude (than the shape) of the distributions.

For completeness, the normalized distributions are also compared to the DSS (normalized) prediction in Fig.~\ref{fig:znorm} (right). As can be seen, the scale dependence is somehow reduced especially at large values of $\zh$, of the order of $30$--$40\%$. Clearly the (normalized) distributions prove really different depending on the FF set used in the calculation. 
This illustrates how the normalized distributions of hadrons inside jets in p--p collisions at the LHC could also bring significant constraints on the current knowledge of fragmentation functions.

Another way to compare the different predictions is to compute the mean value of $\zh$, $\langle\zh\rangle$.  This quantity has the obvious  advantage to also characterize the shape while being insensitive to the magnitude of the distribution $\dd\sigma/\dd\zh$. The numbers corresponding to the various FF sets are given in Table~\ref{table:meanz}, the errors quoted in the case of the DSS FF set corresponds to the scale variation. When looking at $\langle\zh\rangle$ of charged hadrons, the features already discussed are clearly apparent. Namely, the BFGW and DSS (and HKNS within the scale uncertainty) are compatible, $\langle\zh\rangle=0.52\pm0.02$ and $\langle\zh\rangle=0.51$, while significantly smaller values, $\langle\zh\rangle=0.44$, are reported in the case of AKK08 and Kretzer, which FF are much softer. Similar observations can be made for kaon and (anti)proton production, which are discussed in the next section.

\begin{table}[h]
  \centering
  \begin{tabular}{p{3.cm}ccc}
    \hline\hline
    FF set &$\text{h}^++\text{h}^-$& $\text{K}+\overline{\text{K}}$& $\text{p}+\overline{\text{p}}$\\
    \hline\\[-0.2cm]
    DSS& $0.52\pm0.02$& $0.57\pm0.02$& $0.58\pm0.03$ \\[0.3cm]
    AKK08& $0.44$ & $0.49$& $0.44$ \\
    BFGW& $0.51$& -- & --\\
    HKNS& $0.54$ & $0.60$& $0.46$ \\
    Kretzer& $0.44$ & $0.48$& -- \\
  \hline\hline
  \end{tabular}
  \caption{Mean value of  $\langle \zh \rangle$ for $30 < \pth < 200$~GeV and $\ptjet> 30$~GeV. The error indicated in the DSS results reflects the scale uncertainty of $\zh$.}
  \label{table:meanz}
\end{table}

\subsection{Correlations with identified hadrons}\label{se:identifiedhadrons}

As mentioned in the introduction, the most important constraints are put on (quark) fragmentation functions into charged hadrons, due to the the abundance of \epem\ precise measurements. After demonstrating the constraints brought by charged hadron momentum spectra inside jets, we investigate more specifically the production of identified hadrons, kaons and protons, in this section. Calculations are carried out using AKK08, DSS and Kretzer FF sets only (kaon and proton FF are not available in the BFGW parametrization).

\subsubsection{Kaons}\label{se:kaons}

\begin{figure}[htb]
\begin{center}
    \includegraphics[width=14cm]{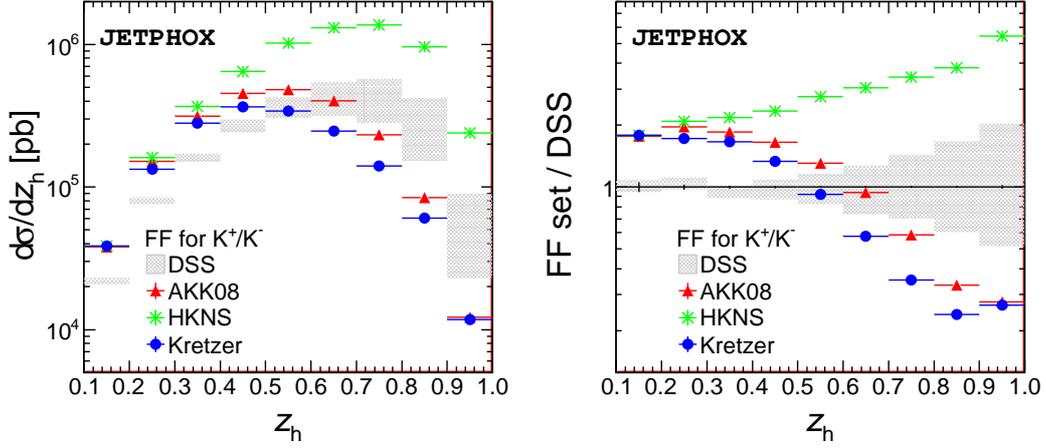}
  \end{center}
\caption{Left: $\zh$ distributions of charged kaons inside jets, using the AKK08, DSS and Kretzer FF sets. The band indicates the scale dependence of the DSS calculation (see text). Right: Same distributions normalized to the DSS prediction.}
  \label{fig:zabsk}
\end{figure}

Using the same cuts as for inclusive charged hadrons, the $\zh$ distributions of kaons inside jets using the different FF sets are shown in Fig.~\ref{fig:zabsk} (left). As can be seen, the differences are very large and significantly beyond the scale uncertainty of the DSS set prediction. This is confirmed in Fig.~\ref{fig:zabsk} (right) where each prediction is normalized to that of DSS. This ratio takes extreme values at high $\zh$: at $\zh=0.8$ it is $r=0.3$ for AKK08 and Kretzer and almost $r=5$ for HKNS. Note also the discrepancy between the various sets and DSS at small values of $\zh \lesssim0.4$.

As mentioned in section~\ref{se:chargedhadrons}, the mean value of the $\zh$ distributions of charged hadrons can be used to discriminate among the various FF sets. It is also the case for kaon production (see Table~\ref{table:meanz}) where a rather large value $\langle \zh \rangle \simeq 0.6$ is reported for DSS and HKNS while $\langle \zh \rangle \simeq 0.5$ for AKK08 and Kretzer fragmentation functions.

\subsubsection{Protons}\label{se:protons}

Finally we discuss in this section the distributions of (anti)protons inside jets. Due to a lack of constraints from data, the fragmentation functions into protons+antiprotons is by far the most uncertain (see e.g. Fig.~\ref{fig:ffgenerator}).

\begin{figure}[htb]
\begin{center}
    \includegraphics[width=14cm]{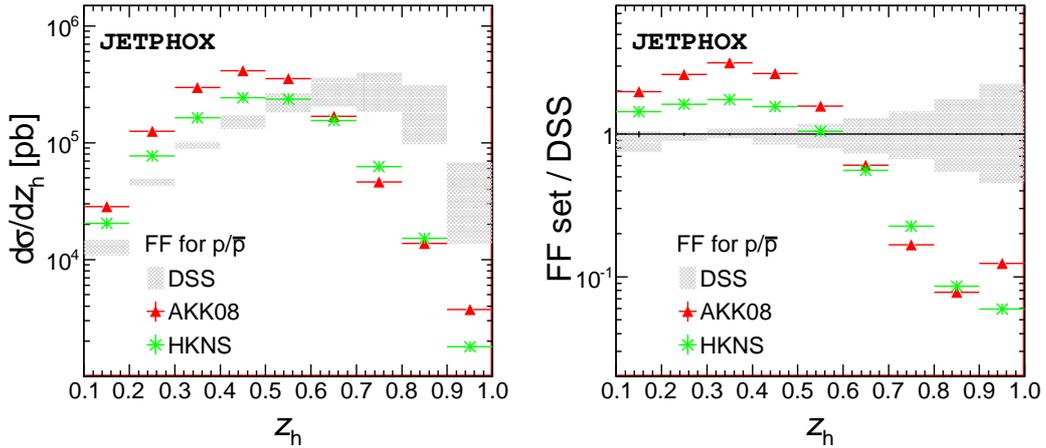}
  \end{center}
\caption{Left: $\zh$ distributions of protons+anti-protons inside jets, using the AKK08, BFGW, DSS and Kretzer FF sets. The band indicates the scale dependence of the DSS calculation (see text). Right: Same distributions normalized to the DSS prediction.}
  \label{fig:zabsp}
\end{figure}

As shown in Fig.~\ref{fig:zabsp}, the $\zh$ distribution of protons+antiprotons inside jets exhibit a very different behavior depending on which fragmentation function set is used, DSS, AKK08 and HKNS (Kretzer set is not available for protons). In this channel, predictions using HKNS prove remarkably similar to those using AKK08 and much softer than the expectations from DSS (this is also true at the level of the FF themselves, Fig.~\ref{fig:ffgenerator}). The ratio between AKK08/HKNS and DSS is  $r=3$ at small $\zh\simeq0.4$ and as low as $r\simeq0.1$ in the largest $\zh$ bins.
 The mean values of $\zh$ reflect also these differences, with $\langle \zh \rangle\simeq 0.45$ for AKK08 and HKNS, and $\langle \zh \rangle\simeq 0.6$ for DSS (see Table~\ref{table:meanz}).
 
\section{Conclusion}\label{se:conclusion}

A NLO perturbative analysis of hadron--jet momentum correlations in p--p collisions at the LHC has been carried out. Results indicate that predictions using various FF sets available exhibit large differences, beyond the scale dependence of the NLO calculation. This is a clear sign that this observable --~which can be measured with a high statistical accuracy at the LHC~-- could be used in order to bring extra constraints, especially in the gluon sector and at large values of $z$ for which the spread of theoretical predictions is largest. This is particularly true regarding kaon and proton spectra inside jets, especially if hadron identification can be performed up to rather large momenta.

\acknowledgments

We thank Patrick Aurenche and David d'Enterria for interesting discussions. This work is funded by ``Agence Nationale de la Recherche'' under grant ANR-PARTONPROP.

\providecommand{\href}[2]{#2}\begingroup\raggedright\endgroup

\ed